# Large colloidal probes for atomic force microscopy: fabrication and calibration issues

M. Chighizola[1], L. Puricelli[1], L. Bellon[2], A. Podestà[1,*]

[1] C.I.Ma.I.Na. and Dipartimento di Fisica "Aldo Pontremoli", Università degli Studi di Milano, via Celoria 16, 20133 Milan, Italy.

[2] Univ. Lyon, ENS de Lyon, Univ. Claude Bernard, CNRS, Laboratoire de Physique, F-69342 Lyon, France.

[*]Corresponding author: alessandro.podesta@mi.infn.it

**ABSTRACT**

Atomic force microscopy (AFM) is a powerful tool to investigate interaction forces at the micro and nanoscale. Cantilever stiffness, dimensions and geometry of the tip can be chosen according to the requirements of the specific application, in terms of spatial resolution and force sensitivity. Colloidal probes (CPs), obtained by attaching a spherical particle to a tipless (TL) cantilever, offer several advantages for accurate force measurements: tunable and well-characterisable radius; higher averaging capabilities (at the expense of spatial resolution) and sensitivity to weak interactions; a well-defined interaction geometry (sphere on flat), which allows accurate and reliable data fitting by means of analytical models. The dynamics of standard AFM probes has been widely investigated, and protocols have been developed for the calibration of the cantilever spring constant. Nevertheless the dynamics of CPs, and in particular of large CPs, with radius well above 10 μm and mass comparable, or larger, than the cantilever mass, is at present still poorly characterised. Here we describe the fabrication and calibration of (large) CPs. We describe and discuss the peculiar dynamical behaviour of CPs, and present an alternative protocol for the accurate calibration of the spring constant.

# 1 INTRODUCTION

Atomic force microscopy (AFM) is a powerful tool to investigate molecular interactions at bio-interfaces, as well as their mechanical properties, with nanometric spatial resolution and 1-10 pN force resolution[1,2].

The force sensitivity of AFM is provided by the use of probes consisting of an elastic microlever (cantilever) with an interacting tip at his apex. A deflection $\Delta z$ of the cantilever as small as 1/10 of a nanometer can be detected by means of an optical beam deflection[3], and translated into a tip-sample force $F$ according to Hook's law: $F = k_{app}\Delta z$. The value of the spring constant of the cantilever can be selected in the range 0.01-100 N/m. Here an apparent spring constant $k_{app}$ is used because due to the cantilever tilt $\theta$ (usually $\theta = 10°-15°$, see Figure 1i), both the force $F$ and the measured deflection $\Delta z$ are perpendicular to the sample surface, rather than to the cantilever axis. The intrinsic spring constant $k$ in turn relates the deflection and the force normal to the cantilever axis.

While sharp AFM tips are mandatory when high spatial resolution is necessary, large spherical probes, also known as colloidal probes (CPs)[4–7], can be advantageous when one or more of the following requirements are important: a well-defined interaction geometry (sphere on flat) and reduced stress and strain in mechanical tests; a good spatial averaging; a broader and smoother surface for probe functionalisation. The well-defined interaction geometry makes CPs a suitable tools for the investigation of mechanical properties of soft or biological samples[8–11]; the larger contact area allows to detect weak interaction forces[12,13] and generally CPs provide an easier interpretation of data in terms of theoretical models[8,11,14,15].

The obvious price to pay when using CPs is the dramatic reduction of the spatial resolution, although the optimal spatial resolution clearly depends on the typical length scale of the physical phenomena under investigation. For example, as long as one is interested in characterizing the average rigidity of mesoscopic cellular regions (cell body vs periphery), tissues or extracellular matrices, CPs provide enough spatial resolution despite their micrometric radii[9,16–18].

Large CPs, where the radius R can be far larger than 10 µm, are very useful to study extremely weak interaction forces, or the mechanics of samples over a wide range of contact area, averaging nanoscale inhomogeneities and local variations of the sample (e.g. in the case of the mechanical response of tissues). Large CPs allows to better compare the results obtained by AFM to those obtained by other indenters[19–22] and in principle to bridge nano- and microscale mechanics to macroscale mechanics[23,24].

Despite the many advantages of CPs, some issues related to the calibration of these probes must be addressed in order to exploit the full potential of these probes, especially in the case of large CPs[25]. In particular, calibration of AFM probes require the determination of the intrinsic and apparent spring constants $k$ and $k_{app}$ and of the static deflection sensitivity $S_z$ (in nm/V), also known as inverse optical lever sensitivity (invOLS), which converts the raw output $\Delta V$ of the photodetector (in Volts) into a deflection $\Delta z$ (in nm) normal to the sample surface: $\Delta z = S_z \Delta V$[26]. The force $F$ is then calculated as:

$$F = k_{app} S_z \Delta V \qquad \text{(Eq. 1)}$$

The calibration of both $k$ ($k_{app}$) and $S_z$ for CPs presents issues, which become more important as the probe radius increases.

The issues related to the calibration of CPs arise when the radius and mass of the attached sphere are comparable to the length and mass of the cantilever, respectively. In order to describe the effect of the attached sphere on the cantilever dynamics, two dimensionless parameters can be introduced (see also the Methods section)[27]: the reduced mass $\tilde{m}$ is the ratio of the mass of the sphere $m_S$ to the mass of the cantilever $m_C$; the reduced gyration radius of the sphere $\tilde{r}$ is proportional to the ratio of the radius $R$ of the sphere to the length $L$ of the cantilever. The basic cantilever calibration parameters, such as the resonant frequency $v_0$, the quality factor $Q$, the deflection sensitivity $S_z$, the spring constant $k$, depends on $\tilde{m}$ and $\tilde{r}$.

In this short report we discuss some critical aspects of CP production and calibration, and present methods to minimise their impact on the accuracy of the AFM measurements.

## 2 METHODS

### 2.1 Fabrication of the colloidal probes

We have produced eight monolithic glass probes following an established custom protocol[28] (see Table 1).

| Probe | Bead Radius (nm) | Cantilever Length (μm) | Cantilever Width (μm) | Cantilever Thickness (μm) | $\tilde{m}$ | $\tilde{r}$ | $\Delta L/L$ | $\beta$ |
|---|---|---|---|---|---|---|---|---|
| CM1 | 2189 | 450 | 50 | 2 | $9.76 \times 10^{-4}$ | $5.7 \times 10^{-3}$ | 0.048 | 0.817 |
| CM2 | 4462 | 450 | 50 | 2 | $8.18 \times 10^{-3}$ | $1.16 \times 10^{-2}$ | 0.053 | 0.826 |
| CM3 | 7696 | 450 | 50 | 2 | $4.24 \times 10^{-2}$ | $2.02 \times 10^{-2}$ | 0 | 0.849 |
| CM4 | 36130 | 125 | 30 | 2 | 7.02 | $9.5 \times 10^{-2}$ | 0.088 | 1.018 |
| TM1 | 1767 | 125 | 30 | 4 | $1.54 \times 10^{-3}$ | $1.67 \times 10^{-2}$ | 0 | 0.817 |
| TM2 | 4292 | 125 | 30 | 4 | $2.21 \times 10^{-2}$ | $4.06 \times 10^{-2}$ | 0.112 | 0.836 |
| TM3 | 7690 | 125 | 30 | 4 | $1.27 \times 10^{-1}$ | $7.2 \times 10^{-2}$ | 0.175 | 0.8768 |
| TM4 | 31752 | 125 | 30 | 4 | 9.62 | $3 \times 10^{-1}$ | 0.501 | 1.111 |

*Table 1. Properties of the CPs fabricated for this study.*

We used borosilicate glass spheres (Thermo Fisher Scientific) with radius $R$ = 2, 5, 10 μm, and soda-lime spheres (Polyscience) with nominal radius $R$= 30 μm; we could not find on the market borosilicate glass spheres with radius above 10 μm. These latter spheres are of far better quality than soda-lime glass spheres, in terms of surface roughness and morphology. Soda-lime glass spheres present extended surface defects (including craters, protrusions) (see Figure 1).

In order to span a wide range of reduced masses and gyration radii we used two different kind of silicon tipless (TL) cantilevers: tapping mode (TM) (TL-NCH-50, Nanosensor) and contact mode (CM) (TL-CONT, Nanosensor).

After mounting the cantilever into the AFM (Bioscope Catalyst, Bruker) tip holder, the beads were captured using the XY motorised stage of the AFM. The AFM is integrated in an optical inverted microscope (Olympus X71), which helps locating and capturing the beads. Depending on the size of the spheres, environmental capillary adhesion or Vaseline is exploited to attach the sphere to the TL cantilever. The cantilevers with the attached beads are then transferred into a high-temperature oven (PXZ series, Fuji Electrics) already at the target temperature, and heated for two hours. We used different temperatures for the different materials of the sphere: 780°C for borosilicate and 700°C for soda-lime glass, respectively. These temperatures are slightly below the softening point of the materials, which is qualitatively defined as the temperature at which a solid object begins to collapse under its own weight. After two hours, the microsphere is covalently attached to the cantilever; the glass locally melts at the sphere-cantilever interface, forming a neck, which is more or less extended depending on curing time and temperature.

Besides being much smoother and less dispersed in radius, borosilicate glass spheres present typically a lower deformation after the heating procedure (Figure 1a,b). Moreover, the radius of soda-lime glass spheres can differ significantly from the nominal one.

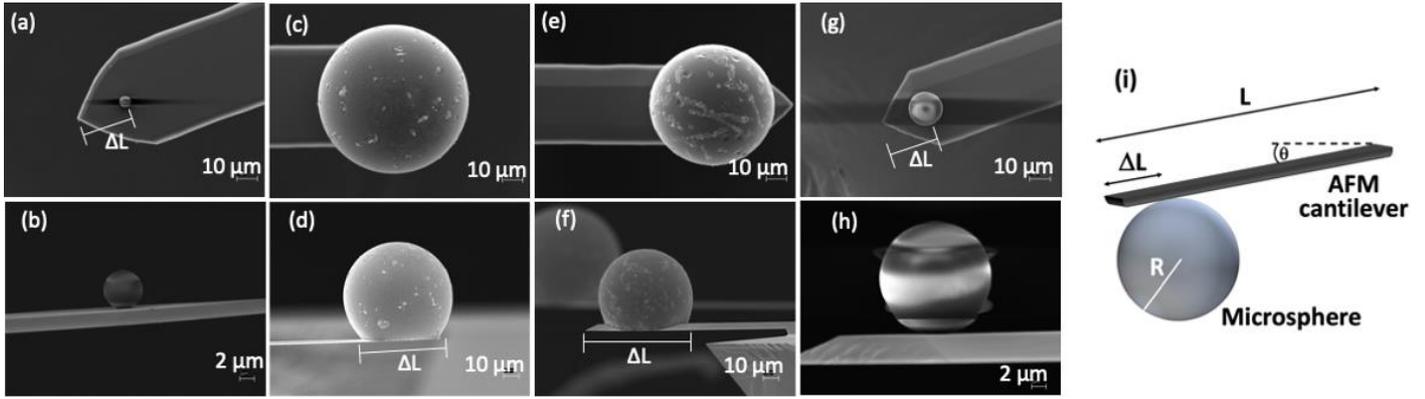

*Figure 1. SEM images of some CPs fabricated (see table 1). (a) CM1, (b) CM1 at higher magnification, (c,d) CM4, (e,f) TM4, (g,h) TM3. (i) Schematics of a CP, with the relevant parameters highlighted.*

## 2.2 Calibration of colloidal probes
### 2.2.1. Radius

1) $R \leq 10$ μm. The characterization of the radius of the CP is obtained by an AFM reverse imaging of the probe, as described in Ref.[28]. The CP is used to scan a spiked calibration grating (TGT1, Tips Nano). Spikes are separated by approximately 3 μm, therefore a 20μm x 20μm image typically contains a hundred independent replicas of the probe apical region. Upon collecting several such images, the volume, projected area, and height of each replica are measured. By fitting a spherical cap model to the experimental data, in the form, for example, of a volume versus height curve: $V = \frac{\pi}{3}h^2(3R - h)$, it is possible to determine the value of the sphere radius $R$, with an accuracy as good as 1%. Since the spikes are relatively close, the sphere can penetrate through them by only a few hundred nanometers, at best. For this reason, as long a gratings with sparser, and taller spikes are not available, the application of this method is limited to relatively small spheres.

2) $R = 30$ μm. CPs with $R > 10$ μm are too large for the calibration by reverse AFM imaging. The radius of these probes is measured optically, using a calibrated metallographic optical microscope equipped with a calibrated 50X objective (AXIO, Zeiss). The error associated to the optical characterisation of the probe radius is 1 μm (standard deviation). Since the nominal radius of the soda-lime spheres attached to the TL cantilevers is $R = 30$ μm, the relative error of its characterisation is approximately 3%.

**2.2.2. Nominal and reduced masses**

We calculated the mass of the spheres $m_s$ using the nominal densities of borosilicate glass $\rho_{BG}$ = 2.23 g/cm$^3$ and soda-lime glass $\rho_{SLG}$ = 2.52 g/cm$^3$ and the calibrated radius $R$ as: $m_s = \rho_{BG/SLG} \frac{4}{3} \pi R^3$.

The mass of the cantilever $m_c$ was evaluated assuming a rectangular section, using the nominal dimensions (length $L$, width $w$, thickness $t$) and the silicon density $\rho_{Si}$ = 2.33 g/cm$^3$ as: $m_c = \rho_{Si} L w t$. We ignored the fact that some cantilevers have a trapezoidal cross section.

The total mass $m_p$ of the probe was evaluated by summing the cantilever and the sphere masses: $m_p = m_c + m_s$.

The reduced mass $\tilde{m}$ and the reduced gyration radius $\tilde{r}$ of the sphere are defined as[27]:

$$\tilde{m} = \frac{m_s}{m_c}, \tilde{r} = \sqrt{\frac{7}{5}} \frac{R}{L} \qquad (2)$$

We evaluated the effective mass $m^*$ of the cantilever, before and after the attachment of the sphere, as the effective mass of a SHO, using the standard formula: $m^* = k/\omega_0^2$, where $k$ is the spring constant and $\omega_0$ is the angular frequency of the oscillator.

**2.2.3. Deflection sensitivity**

The calibration of $S_z$ is performed by engaging the AFM tip on a very rigid surface and collecting a series of raw deflection vs z-piezo displacement curves. Assuming that neither the tip nor the surface are deformed, the deflection sensitivity $S_z$ is calculated as the inverse of the slope of the curves[26]. This sensitivity is referred to as the static deflection sensitivity, and differs from the dynamic deflection sensitivity $S_{z,dyn}$[29,30]; the latter sensitivity must be used when the cantilever oscillates instead of being statically bent due to an end loading, as in the application of the thermal noise method for the calibration of the cantilever spring constant.

It must be noted that $S_z$ strongly depends on the loading configuration[31–33]. In particular, the static sensitivity of a tilted tipless cantilever differs from that of a standard tilted cantilever, with a sharp tip, or a sphere, attached to its end.

In the case of a standard probe, the force is not applied at the cantilever end, but rather at the tip apex, which is some distance apart, therefore a torque is generated. Moreover, because of the tilt angle $\theta$, during the standard operation of an AFM the interaction force is perpendicular to the sample surface, instead of the cantilever axis, while during the thermally driven oscillation of the cantilever the effective driving force is always perpendicular to the cantilever axis. The corresponding cantilever displacement is therefore different from that obtained by the application of the same force at the cantilever end.

Corrections for the deflection sensitivity $S_z$ are necessary, to account for the differences in the loading configuration of tipless and tipped cantilevers in the different experimental conditions. Correction factors are therefore used to determine both the intrinsic spring constant $k$ using the thermal noise method, and the apparent spring constant $k_{app}$ used to rescale the raw deflection signal into a force (Eq. 1). While the latter correction factor is discussed in section 2.2.5, the first factor is introduced in the following paragraphs.

As mentioned, during the thermal noise calibration, the cantilever is loaded by an effective thermal force located at the cantilever end and acting perpendicularly to the cantilever axis. Assuming that the deflection sensitivity depends on the parameters $\theta$, $R$, $\Delta L$ introduced in Figure 1i, the correction factor is therefore $\hat{s} = S_z^{0,0,0}/S_z^{\theta,R,\Delta L}$, where $S_z^{\theta,R,\Delta L}$ represents the deflection sensitivity measured in a generic loading configuration and $S_z^{0,0,0}$ is the sensitivity in the thermal loading configuration. According to Ref.[32] (Eq. 3 and other equations therein):

$$1/\hat{s}^2 = \left(1 - \frac{\Delta L}{L}\right)^2 \left[\frac{1 - \frac{3}{2}\frac{R/L}{\left(1-\frac{\Delta L}{L}\right)}\tan\theta}{1 - 2\frac{R/L}{\left(1-\frac{\Delta L}{L}\right)}\tan\theta}\right]^2 \cos^2(\theta) \tag{3}$$

The $\cos^2(\theta)$ term in Eq. 3 accounts for the tilt $\theta$ of the cantilever[34]; the term in square brackets accounts for the torque applied to the cantilever because of the finite height of the tip; the first term *(1 - ΔL/L)²* accounts for the fact that when the static deflection sensitivity is calibrated, the deflection at the loading point, rather than at the end of the cantilever, is measured[32,33]. A special case of Eq. 3 was first proposed by Hutter[33] under the hypothesis *ΔL = 0;* in this case, Eq. 3 simplifies to:

$$1/\hat{s}^2 = \left[\frac{1 - \frac{3R}{2L}\tan(\theta)}{1 - 2\frac{R}{L}\tan(\theta)}\right]^2 \cos^2(\theta) \tag{4}$$

For standard AFM tips, the radius $R$ of the sphere in Eqs. 3,4 must be replaced by the tip height $H$[32].

### 2.2.4. Spring constant and thermal noise and CPs

Several approaches has been developed to characterize the spring constant of a cantilever[6,35–40], among which, the thermal noise method[35,36,41,42] is based on the equipartition theorem.

Under the assumption that the mass $m_t$ of the tip is negligible with respect to the mass $m_c$ of the cantilever, the probe (cantilever + tip) can be modelled as a point body with effective mass $m^*$, bound to an ideal spring with stiffness $k$, in thermal equilibrium with its environment.

Assuming that the tip oscillates only along the direction perpendicular to the cantilever, and being $z_\perp$ its displacement relative to the equilibrium position, its average potential energy is $U = \tfrac{1}{2}k<z_\perp^2>$, where $\sqrt{<z_\perp^2>}$ is the thermally driven root mean square oscillation amplitude of the free end of the cantilever, where the tip is located, ranging from approximately 5 to 500 pm. According to the equipartition theorem, $U = \tfrac{1}{2}k_B T$, where $k_B$ is the Boltzmann constant and $T$ is the local absolute temperature. It then follows:

$$\tfrac{1}{2} k <z_\perp^2> = \tfrac{1}{2} k_B T \qquad (5)$$

If $<z_\perp^2>$ is measured, the cantilever spring constant $k$ is then calculated from Eq. 5 as:

$$k = \frac{k_B T}{<z_\perp^2>} \qquad (6)$$

When the optical beam deflection method is used to measure the cantilever deflection, with the laser aligned at the end of the cantilever, Eq. 6 is replaced by Eq. 7[32,33,43]:

$$k = \frac{\hat{\alpha}}{\chi^2} \frac{k_B T}{(\hat{s} S_z)^2 P_V} \qquad (7)$$

In Eq. 7, $P_V = <\Delta V^2>$ is the area below the first resonant peak in the power spectral density (PSD) of the raw deflection signal (in units of V²/Hz).

The factor $\hat{\alpha}$ in Eq. 7 takes into account the fraction of the total vibrational energy stored in the first normal mode of the cantilever, when one goes beyond the single harmonic oscillator (SHO) approximation[41]. For the first normal mode of a TL rectangular cantilever, $\hat{\alpha} \equiv \hat{\alpha}_1 = 0.971$.

The factor $\chi$[29,30,43] corrects for the differences between the static (which is typically measured) and dynamic deflection sensitivities: $\chi = S_{z,dyn}/S_z$[41,44]. Indeed, as pointed out by Butt and Jaschke in their seminal work[41], the cantilever inclination, rather than its vertical displacement, is measured using the optical beam deflection apparatus; the slope of the free end of the statically loaded cantilever differs from those of the modal shapes of the oscillating cantilever[41,43,45]. For the first normal mode of a TL rectangular cantilever, assuming a negligibly small laser spot aligned at the end of the cantilever, $\chi \equiv \chi_1 = 1.09$. For the general case, the reader is referred to Refs[32,45] for the calculation of $\chi$.

The factor $\hat{s}$ represents the correction to the deflection sensitivity due to the change in the loading condition of the cantilever (Eq. 3,4).

In principle, the thermal noise method could be applied to higher normal modes, in which case Eq. 7 should be rewritten in terms of the parameters $P_{V,i}$, $\hat{\alpha}_i$, $\chi_i$, where the index $i = 1,2,3,...$ refers to the $i$-th normal mode. In practice, due to the finite laser spot size, which causes an averaging of different cantilever inclinations across a finite area, the first normal mode has to be preferred[30]; moreover, the first normal mode typically provides the largest oscillation amplitude, i.e. the largest signal to noise ratio.

The factors $\hat{\alpha}$ and $1/\chi^2$ can be merged in a single parameter $\beta = \hat{\alpha}/\chi^2$. For the first normal mode of an ideal rectangular cantilevers, $\beta = \beta_0 = 0.817$.

The implementation of the thermal noise model for AFM cantilevers assumes that the mass of the cantilever is uniformly distributed. For this statement to be valid, the mass of the tip must be negligible compared to the mass of the cantilever. When large spheres are attached at the end of a cantilever, this hypothesis is not always satisfied. Since the mass of the microsphere scales up with the cube of radius $R$, it can easily reach values comparable to the mass of the cantilever. This typically happens for a radius $R > 10\mu m$, for a 200 μm long, silicon nitride cantilever, and a glass sphere.

### 2.2.5. Loading point and spring constant

For TL cantilevers, the loading point is located at the free end of the lever. For sharp tips, the loading point corresponds to the location of the tip apex along the cantilever. For CPs, the loading point is located at the sphere apex, i.e. approximately a distance ΔL from the cantilever end (Figure 1i).

The attachment of the sphere to the TL cantilever is not expected to change the spring constant of the probe. Since the determination of the deflection sensitivity $S_z$ can be critical for CPs (see discussion in Section 3.6), it can be convenient to calibrate the intrinsic spring constant $k_{TL}$ of the TL cantilever, before the attachment of the sphere, and then correct its value to determine the intrinsic spring constant $k_{TL}^{LP}$ of the TL cantilever at the loading point of the CP; the latter spring constant can be supposed to be equal to the spring constant of the CP at the loading point, i.e. $k_{TL}^{LP} = k_{CP}^{LP}$.

The spring constant $k_{TL}^{LP}$ of the TL cantilever corresponding to the loading point of the CP is calculated as[46,47]:

$$k_{TL}^{LP} = \left(\frac{L}{L-\Delta L}\right)^3 k_{TL} \tag{8}$$

In Eq. 8, *ΔL* is the distance of the loading point of the CP from the cantilever free end. In the case of rectangular, axisymmetric and relatively long cantilevers, Eq. 8 is valid for arbitrary offset *ΔL*, not only in the limit $\Delta L/L \ll 1$ [47].

Alternatively, the thermal noise method can be applied directly to the CP, provided the measured spring constant $k_{CP}$ is corrected first by Eq. 8 (to obtain $k_{CP}^{LP}$), and then by Eq. 11, to take into account the peculiar dynamics of the CP, as described in the Results.

When the optical beam deflection method is used to measure the cantilever deflection, in order to rescale deflections into forces perpendicular to the sample surface, an apparent spring constant $k_{app}$ must be used, instead of the intrinsic spring constant $k$ [32]. The relation between the apparent and the intrinsic spring constant is:

$$k_{app} = \left\{\left[1 - \frac{3R}{2(L-\Delta L)}\frac{\sin(2\alpha_0+\theta)}{\cos(\theta)}\right]\cos^2(\theta)\right\}^{-1} k \tag{9}$$

where the correction factor accounts for the tilt of the cantilever *θ* and the torque applied on the cantilever because of the finite tip height; it was proposed by Heim et al.[35], and later corrected by Hutter[33] and generalised by Edwards et al.[32] and Wang[48]. For standard AFM tips, the radius *R* of the sphere in Eq. 9 must be replaced by the tip height (and not by half that value), as reported in Ref.[32].

For CPs obtained from TL cantilevers, Eq. 8 and 9 can be combined as:

$$\begin{aligned} k_{CP,app}^{LP} &= \left\{\left[1 - \frac{3R/L}{2(1-\Delta L/L)}\tan\theta\right]\cos^2(\theta)\right\}^{-1}\left(\frac{L}{L-\Delta L}\right)^3 k_{TL} \\ &= \left\{\left[1 - \frac{3R/L}{2(1-\Delta L/L)}\tan\theta\right]\cos^2(\theta)\right\}^{-1} k_{CP}^{LP} \end{aligned} \tag{10}$$

Eq. 10 allows to calculate the apparent spring constant, at the loading point, of the CP from the intrinsic spring constant of the previously calibrated TL cantilever.

The intrinsic spring constants are calibrated using the thermal noise method[36,41]. Some details of the implementation of the thermal noise method for CPs are discussed in the Results section. The cantilever deflection signal is extracted from the AFM using a signal access module (Intermodulation Products) and sampled using an external ADC board (NI-DAQ PCI-6115). Signal acquisition and processing is performed using custom software (LabView, National Instruments) (see Ref.[43] for practical indications). Alternatively, the Sader method can be used to calibrate the intrinsic spring constant of the cantilevers[40,49].

# 3 RESULTS

## 3.1. Impact of the added mass on the dynamics of CPs

The Laser Doppler Vibrometry (LDV) and in general interferometric techniques[42,50–52] permit to measure with great accuracy the cantilever deflection instead of inclination, also in the case of oscillating cantilevers, including for the calibration of the spring constant by the thermal noise method. Interferometric techniques offer the advantage of not requiring the calibration of deflection sensitivity.

Using interferometric approaches it was recently demonstrated that the cantilever oscillation dynamics changes if a mass is attached to the free end[27,53]. In particular Laurent et al.[27] have measured the shape of first five eigenmodes along the profile of a rectangular cantilever, as a function of the reduced mass of the sphere $\tilde{m}$, and of the reduced gyration radius of the sphere $\tilde{r}$, introduced in the Methods. Apart from causing a decrease of the resonant frequency, the large attached mass of the sphere induces a shift of the normal mode nodes towards the free end of the cantilever, with the progressive shift of the centre of mass towards the cantilever free end and the storage of a greater fraction of vibrational energy in the first normal mode. As a consequence, both factors $\hat{\alpha}$ and $\chi$ are expected to be influenced by the presence of a large sphere at the end of the cantilever. It turned out indeed that the factor $\beta_0$ calculated for an ideal TL cantilever is no longer accurate, and leads to an underestimation of the spring constant measured by the thermal noise method[27].

Figure 2 shows the values of the factor $\beta$ calculated according to the procedure described in Ref.[27] based on the modified modal shape functions for a sphere-loaded cantilever, for both $\tilde{m}$ and $\tilde{r}$ varying over a broad range, considerably extending the previously reported results.

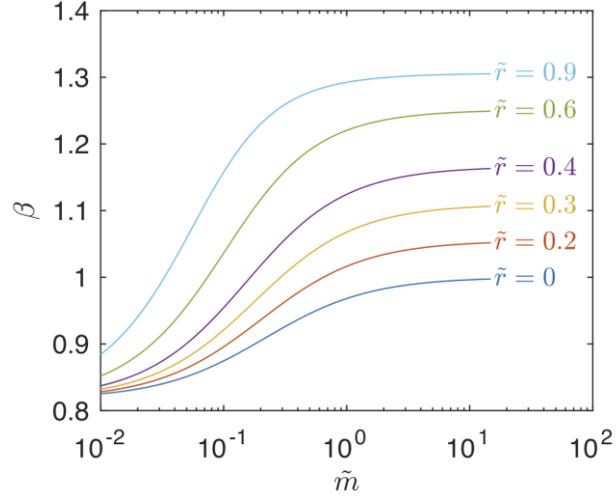

*Figure 2. Scaling of the factor β as function of $\tilde{m}$, for different $\tilde{r}$.*

Once the spring constant $k_{CP,0}$ of the CP has been calibrated by the thermal noise method using the standard prefactor $\beta_0$, the correct spring constant $k_{CP}$ is calculated according to the formula:

$$k_{CP} = \frac{\beta}{\beta_0} k_{CP,0} \tag{11}$$

Representative curves $\beta \equiv \beta(\tilde{m})$ for different values of $\tilde{r}$ are reported in Figure 2. Table 2 reports the values of the correction factor $\beta$ for different values of $\tilde{m}$ and $\tilde{r}$.

|  | | $\tilde{r}$ | | | | | | | | | |
|---|---|---|---|---|---|---|---|---|---|---|---|
|  | | 0 | 0,1 | 0,2 | 0,3 | 0,4 | 0,5 | 0,6 | 0,7 | 0,8 | 0,9 | 1 |
| $\tilde{m}$ | 0 | 0,817 | 0,817 | 0,817 | 0,817 | 0,817 | 0,817 | 0,817 | 0,817 | 0,817 | 0,817 | 0,817 |
|  | 0,001 | 0,818 | 0,818 | 0,818 | 0,819 | 0,819 | 0,820 | 0,821 | 0,822 | 0,823 | 0,824 | 0,826 |
|  | 0,002 | 0,819 | 0,819 | 0,820 | 0,820 | 0,821 | 0,823 | 0,825 | 0,827 | 0,829 | 0,832 | 0,834 |
|  | 0,005 | 0,821 | 0,822 | 0,823 | 0,825 | 0,827 | 0,831 | 0,835 | 0,840 | 0,846 | 0,852 | 0,859 |
|  | 0,01 | 0,825 | 0,826 | 0,828 | 0,832 | 0,837 | 0,844 | 0,852 | 0,861 | 0,872 | 0,884 | 0,897 |
|  | 0,02 | 0,832 | 0,834 | 0,838 | 0,845 | 0,855 | 0,868 | 0,883 | 0,900 | 0,920 | 0,941 | 0,963 |
|  | 0,05 | 0,851 | 0,854 | 0,864 | 0,879 | 0,901 | 0,927 | 0,957 | 0,990 | 1,024 | 1,059 | 1,093 |
|  | 0,1 | 0,875 | 0,880 | 0,896 | 0,921 | 0,954 | 0,994 | 1,036 | 1,079 | 1,121 | 1,158 | 1,190 |
|  | 0,2 | 0,905 | 0,913 | 0,936 | 0,972 | 1,018 | 1,068 | 1,117 | 1,161 | 1,199 | 1,230 | 1,254 |
|  | 0,5 | 0,945 | 0,956 | 0,988 | 1,035 | 1,090 | 1,144 | 1,191 | 1,228 | 1,257 | 1,278 | 1,292 |
|  | 1 | 0,968 | 0,981 | 1,016 | 1,068 | 1,124 | 1,177 | 1,220 | 1,253 | 1,276 | 1,292 | 1,304 |
|  | 2 | 0,983 | 0,996 | 1,034 | 1,087 | 1,144 | 1,195 | 1,236 | 1,265 | 1,285 | 1,299 | 1,309 |
|  | 5 | 0,993 | 1,007 | 1,046 | 1,100 | 1,157 | 1,207 | 1,245 | 1,272 | 1,291 | 1,303 | 1,312 |
|  | 10 | 0,996 | 1,011 | 1,050 | 1,105 | 1,162 | 1,211 | 1,248 | 1,275 | 1,293 | 1,305 | 1,313 |

*Table 2. Values of the correction factor $\beta$ (Eq. 11) for different values of $\tilde{m}$ and $\tilde{r}$.*

**3.2 A survey of CPs from the literature and the present study**

CPs are fabricated with different cantilever and microsphere relative dimensions and masses. Figure 3 reports the results of a literature survey about the values of $\tilde{m}$ and $\tilde{r}$ of CPs used in the experiments. Most of the CPs have relatively small spheres attached to the cantilevers, resulting in a reduced mass $\tilde{m}$ on the order of 0.01 or even below[10,54–56]. Nevertheless, very large CPs are also used[10,57–60], with $\tilde{m}$ values as large as 5, up to 10.

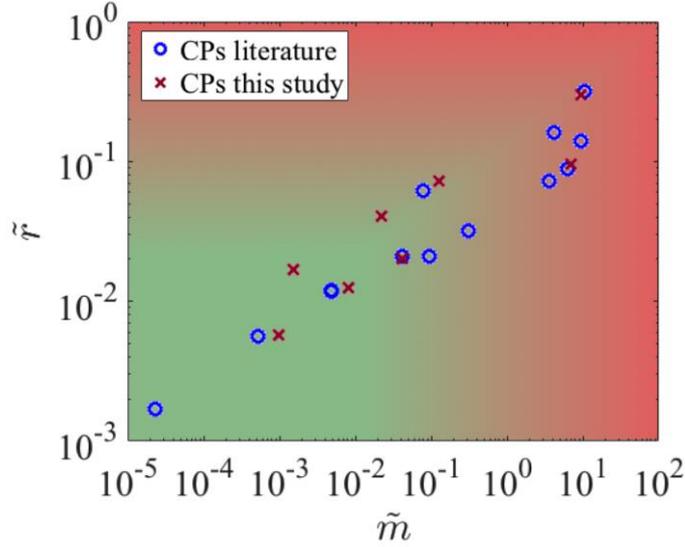

*Figure 3. CPs used by the authors of publications as from a literature survey (blue circles) and CPs fabricated during this study (red crosses) represented according to their reduced mass $\tilde{m}$ and gyration radius $\tilde{r}$. The background colour highlights the region where corrections to the spring constant according to Eq. 11 are negligible (green), and the region where the influence of the added mass becomes relevant (deviation larger than approximately 10%).*

In order to study how $\tilde{m}$ and $\tilde{r}$ of CPs affect the cantilever dynamics, we fabricated several CPs with different dimensions in order to cover the whole range shown in Figure 3. In particular, we produced eight probes, using both shorter and rigid tapping mode (TM), and longer and soft contact mode (CM) cantilevers. The properties of the CPs produced for this study are reported in Table 1 and also in Figure 3 (red crosses). We characterized the spring constant and the dynamics of our probes before and after the attachment of the microsphere.

**3.3 Shift of the resonant frequency of CPs**

Figure 4 shows the thermal noise spectrum, i.e. the raw PSD $P_V$ in units of V$^2$/Hz, of the first resonant peak of the original TL cantilevers, and of the CPs obtained upon attachment of the glass bead. Not surprisingly, as a consequence of the added mass, we observe a decrease of the resonant frequency. This results confirms that a shift in the resonant frequency $v_0$ is present even if the reduced mass is as small as $\tilde{m} = 9.76 \cdot 10^{-4}$ (Figure 5). This drop of the resonant frequency reaches a factor of 5 for largest masses (Figure 4d,h). These results are in agreement with previous works[61,62].

A sharpening of the peak and an increase of the oscillation amplitude are associated to the shift of the resonant peaks towards lower frequencies. For large CPs (Figure 4d,h), the peak is 30 times higher and 20 times narrower compared to the corresponding TL cantilever case. These observations may be somewhat unexpected for probes whose mass is significantly increased. For instance, a reduction of the oscillation amplitude and a widening of the resonant peak is typically observed when the thermal noise spectrum is acquired in liquid, another case where the effective mass of the cantilever is increased significantly, due to the liquid viscosity that causes the cantilever to drag a finite volume of liquid during oscillation. As we will discuss later, this observation can be explained in terms of a combination of effects due to the increase of the inertial mass of probe[62] and the storage of a greater fraction of potential energy in the first mode[27].

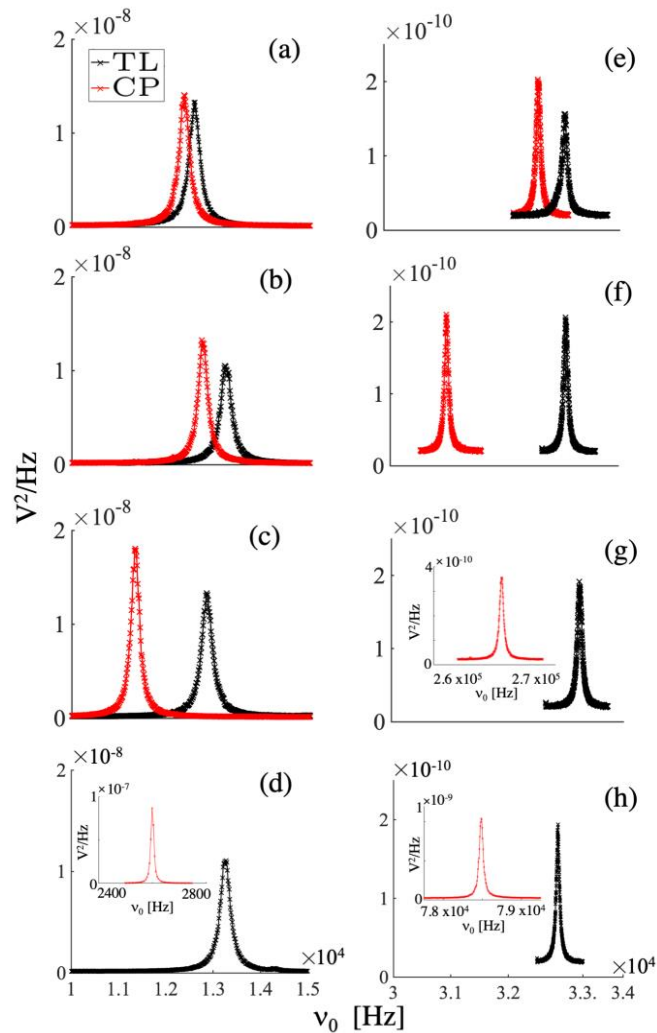

*Figure 4. The PSD $P_V$ at the first normal mode of the cantilever before (black) and after (red) the attachment of the microsphere for contact mode CPs. (a-d) CMs, (e-h) TMs. The mass of the attached sphere increases from top (CM1(a)-TM1(e)) to bottom (CM4(d)-TM4(h)).*

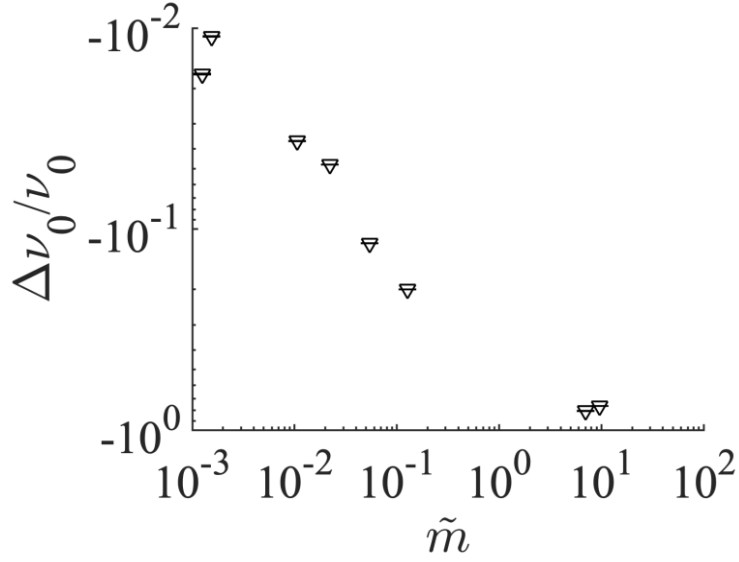

*Figure 5. The relative shift of the resonant frequency $v_0$ versus the reduced mass $\tilde{m}$. As shown in Figure 4, $v_0$ decreases as $\tilde{m}$ increases.*

**3.4 Quality factor**

The quality factor $Q$ represents the ratio between the energy stored in the oscillation and the energy dissipated per cycle because of the viscous damping. $Q$ is defined as:

$$Q = \frac{m^*\omega_0}{\gamma} = 2\pi\frac{m^*v_0}{\gamma} \tag{12}$$

In Eq. 12, $m^*$ is the (effective) mass of the oscillator, $\omega_0 = 2\pi v_0$, and $\gamma$ is the damping coefficient, i.e. the proportionality factor between the tip velocity and the viscous force. $Q$ is related to the width of the resonant peak: the higher is $Q$, the narrower is the peak.

Figure 6 shows that the value of $Q$ of CPs increases, rather than decreasing, with respect to the TL cantilevers, the relative increase being larger for higher values of $\tilde{m}$.

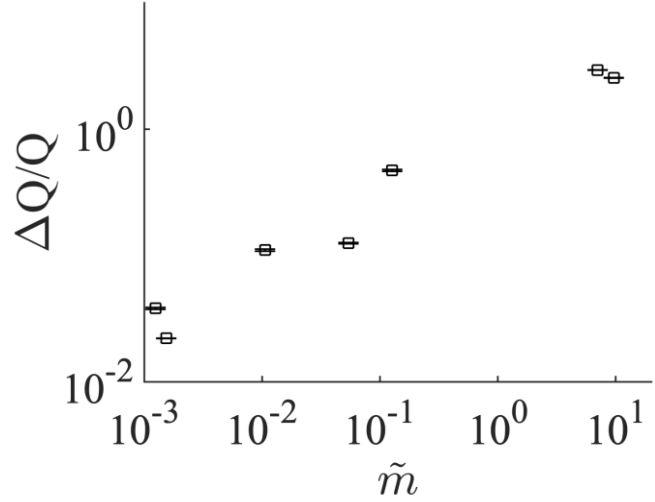

*Figure 6. Relative shift of the quality factor Q versus the reduced mass $\tilde{m}$.*

To better understand the behaviour of $Q$ upon the attachment of the sphere to the TL cantilever, we calculated the effective mass $m^* = k/\omega_0^2$ for each probe, before and after the attachment of the bead. The ratio between the measured effective mass $m^*$ and the nominal mass of the probe $m_p$ (for both TL probes and CPs) is shown in Figure 7a. The measured ratio $m^*/m_p$ for the TL cantilevers is 0.24 ± 0.02 (mean + standard deviation of the mean), in agreement with the theoretical value of 0.25 for the rectangular cantilever[37]. A similar value is measured for those CPs with the smallest reduced mass, where the contribution of the cantilever mass is dominant (see figure 7a). As $\tilde{m}$ increases, the effective mass $m_c^*$ becomes larger than $0.25 m_c$. For large CPs, $m^*$ becomes similar to $m_p$, which is in turn very similar to $m_s$, the mass of the sphere; in this limit, the mass of the sphere is dominant.

Eq. 13 represents a simple model for the effective mass $m_{CP}^*$ of a CP; the concentrated mass of the sphere, located at the free end of the cantilever, is added to the effective mass of the TL cantilever[26]:

$$m_{CP}^* = m_c^* + m_s \qquad (13)$$

Eq. 13 clearly highlights that the cantilever and the sphere contribute differently to the inertia of the probe. The cantilever mass, which is uniformly distributed along the cantilever length, is strongly underestimated in the sum, being weighted by a factor of ¼; the mass of the sphere, instead, being truly concentrated at the free end of the cantilever, is fully represented with weight 1. Figure 7b shows the correlation between the measured $m^*$ and the values of the effective mass predicted by

Eq. 13. The slope of the experimental curve in Figure 7b is 1.052 ± 0.005. In the case of the first six probes (data highlighted in Figure 7c), a better correlation is found, the slope being 1.027 ± 0.008.

Figure 7b shows that Eq. 13 represents a good model for the effective mass of CPs, which in turn confirms the accuracy of the description of (large) CPs as effective SHOs with the total mass concentrated at the loading point. Indeed, as the radius $R$ of the sphere increases, the mass of the sphere increases rapidly, scaling as $R^3$, and the CP becomes more and more equivalent to an SHO with a concentrated mass equal to the sphere mass. Deviations for the largest CPs can be attributed to uncertainties in the determination of the sphere mass, due both to the uncertainties in the sphere density or in the sphere volume. SEM images (Figure 1) show indeed that soda-lime glass spheres possess many defects, which could also be present in the inner volume, as voids, or segregated phases with different density. Moreover, upon softening in the oven, the spheres deform to some extent in the region close to the cantilever. This process is supposed to conserve the volume of the sphere, but could alter the measurement of the radius, which is inferred from the transversal section of the sphere; as a consequence, errors in the volume estimation are likely.

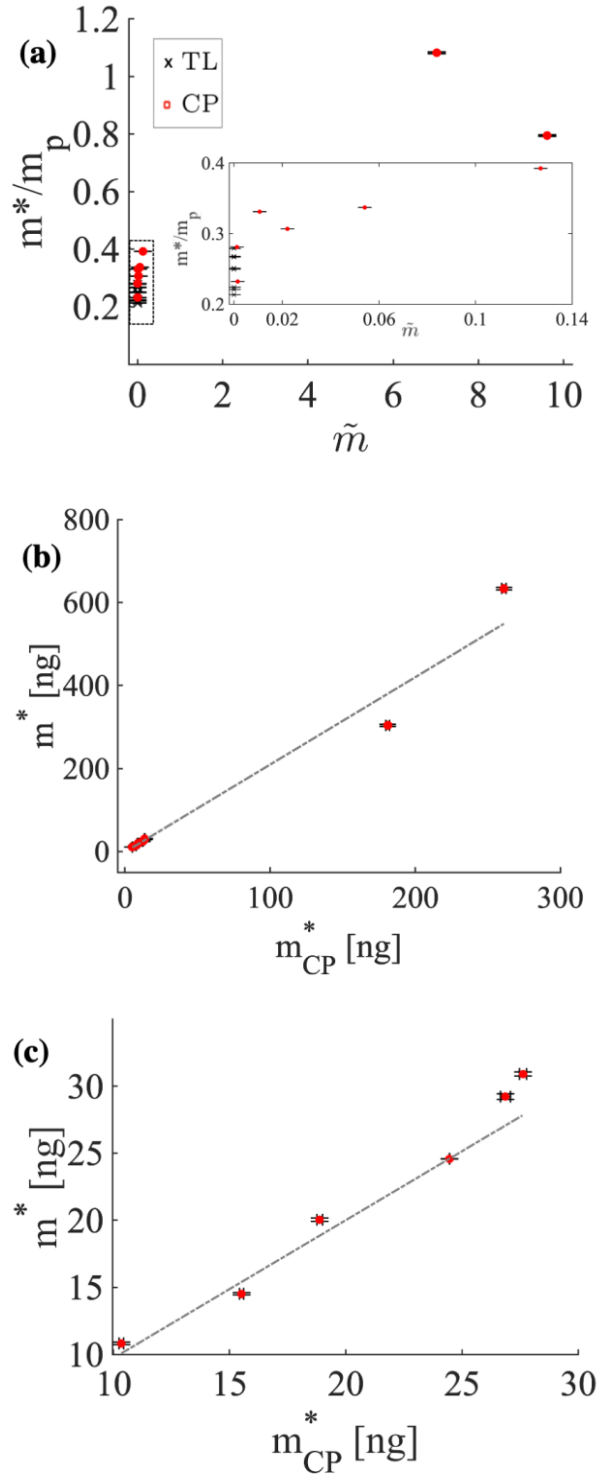

*Figure 7. (a) Ratio of the effective mass $m^*$ to the nominal mass of the probe $m_p$ as function of the reduced mass $\tilde{m}$ for TL cantilevers (black) and CPs (red). (b) Effective mass $m^*$ of the CPs as a function of $m^*_{CP}$ with linear fit (y = mx + b, m = 1.052 ± 0.005, b = -1.14x10$^{-12}$ ± 1.27x10$^{-13}$). (c) Zoom on the data presented in (b) for $m^*_{CP}$ < 30 ng, with their specific linear fit (y = mx + b, m = 1.027 ± 0.008, b = -1.14x10$^{-12}$ ± 5.37x10$^{-13}$).*

The increase of the quality factor is in turn explained by Eq. 12. $Q$ is proportional indeed to the resonant frequency $\nu_0$ and the effective mass $m^*$. Since $\nu_0$ drops by a factor of 5 (Figure 5), while $m^*$ increases by a factor of 40 (it passes from $0.24m_c$ to approximately $10m_c$), the latter change dominates and $Q$ increases (considerably).

Eventually, Figure 7b-c and Figure 6 explain why the amplitude of the resonant peak of the first normal mode is enhanced, as shown in Figure 4. Indeed, large CPs are better described by SHOs, with a significantly larger fraction of vibrational energy stored in the first normal mode; as the quality factor Q increases significantly, the constraint of energy conservation causes the amplitude to increase.

In conclusion, Eq. 12,13 and the observed trends of both resonant frequency and quality factor explain the peculiarity of the thermal noise spectra of CPs, in particular the fact that these probes, despite a marked reduction of the resonant frequency, have extremely sharp resonant peaks, with large $Q$ and amplitude. The apparent anomaly of CPs can be attributed to the fact that, in the AFM community, an increase of the effective mass and a drop in the resonant frequency of the cantilever are typically observed when operating a soft cantilever in liquid[63–65], and they are associated to a marked drop of the quality factor $Q$. With CPs, however, the added mass mechanism is not related to viscous dissipation, therefore in Eq. 12 the term $\gamma$ does not change significantly (at least in air)[62]; moreover, the mass does not change so significantly as for CPs, therefore this change is better balanced by the decrease in the resonant frequency.

**3.5 Spring constant of CPs**

Figure 8a shows a comparison of the spring constants $k_{TL}^{LP}$, calibrated for the TL cantilevers before the attachment of the glass beads, and then corrected for the shift of the loading point (Eq. 8), with the spring constants $k_{CP}^{LP}$ of the CPs calibrated after the attachment of the spheres, corrected for the loading point shift, and eventually for the added mass effect (Eq. 11). A linear regression of the data provides a slope of $1.0090 \pm 0.0006$ confirms the effectiveness of the correction procedures.

Figure 8b,c shows the relative difference between the spring constant $k_{TL}^{LP}$ and the spring constant $k_{CP}^{LP}$ as a function of $\tilde{m}$ and $\tilde{r}$, respectively. The spring constant values are shown before (red), and after (black) the $\beta$-factor correction according to Eq. 11.

As shown in Ref[27], the effect of the added concentrated mass of the sphere is to induce an underestimation of the spring constant, if the standard thermal noise formula (Eq. 7) is used, with $\beta = \beta_0 = 0.817$. When $\tilde{m}$ and $\tilde{r}$ are larger than approximately 0.1, the spring constant starts to decrease, reaching a maximum relative deviation from $k_{TL}^{LP}$ of about 25% (Figure 8b,c, red circles), these deviations are effectively corrected by Eq. 11 (Figure 8b,c, black circles). For $\widetilde{m} < 0.1$, the correction is not very relevant, and falls below the typical noise level.

The small deviations from the $k_{TL}^{LP} = k_{CP}^{LP}$ line reported in Figure 8 could be due, among other factors, to errors in the determination of the deflection sensitivity of the CPs, as discussed below in Section 3.6, in the determination of the loading point, as well as in the possible rigidification effect of the cantilever, due to the attachment of the sphere, which could influence the dynamics of the last part of the cantilever.

### 3.6 Deflection sensitivity and CPs

When big spheres are attached to the cantilever, the standard approach for the calibration of the deflection sensitivity $S_z$ (described in the Methods) can be inaccurate, due to phenomena occurring during the loading phase of the force curve.

In particular, the cantilever tilt angle $\theta$ towards the surface may determine a tangential force component, which is due to friction, between the sphere and the substrate. As in the case of normal forces[32,33], also tangential friction forces applied at the loading point of the cantilever induce a torque[48,60,61,66–68]; this torque in turn induces a modification of the cantilever curvature that impacts on the measurement of $S_z$[67–69].

The friction-related effect is stronger in the high-friction limit, when the probe is pinned on the surface instead of sliding[70]. The friction, and therefore the torque, are stronger when the effective tip height is larger, as in the case of large CPs, and when the normal force is higher, as in the case stiffer cantilevers are used. Noticeably, for mechanical measurements stiffer cantilevers can be used, in combination with large spheres, in order to achieve reasonably high indentations.

Since the AFM optical lever detection scheme is sensitive to the inclination of the cantilever rather than to its deflection, it turns out that the measurement of $S_z$ can be particularly inaccurate for large CPs, especially in conditions of strong adhesion and friction (as in air). In addition, larger adhesion related to the larger radius of CPs in general makes the acquisition of force curves in air, but also in liquid, critical[71].

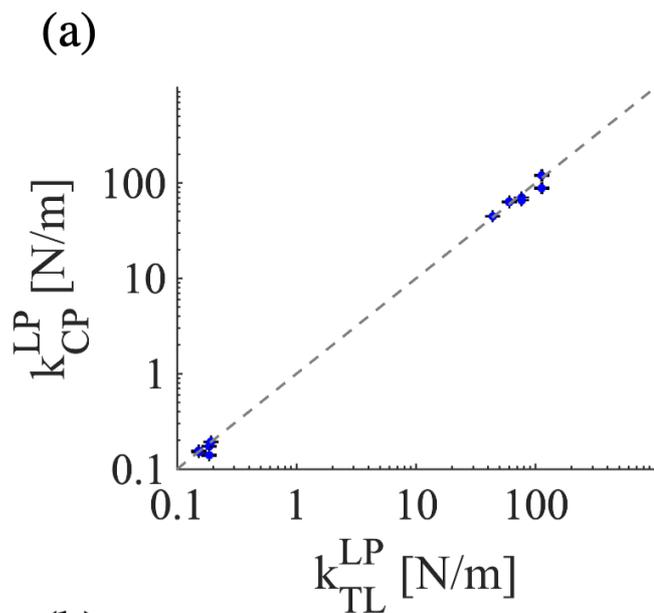

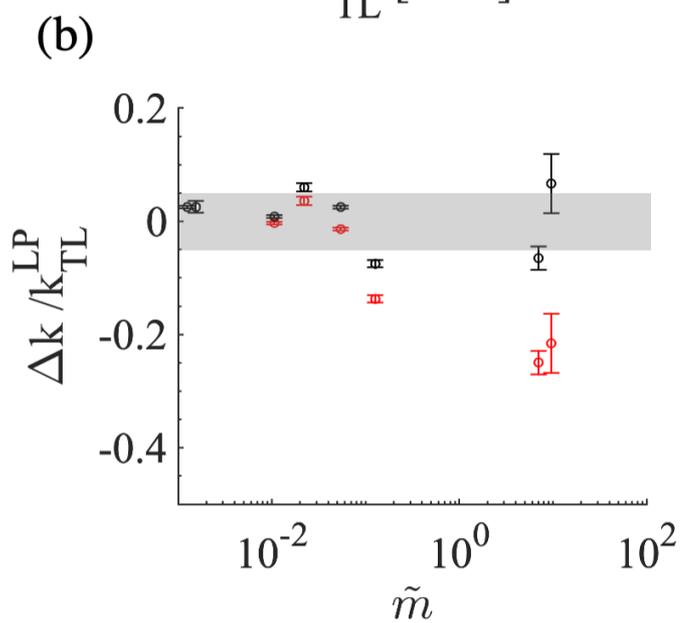

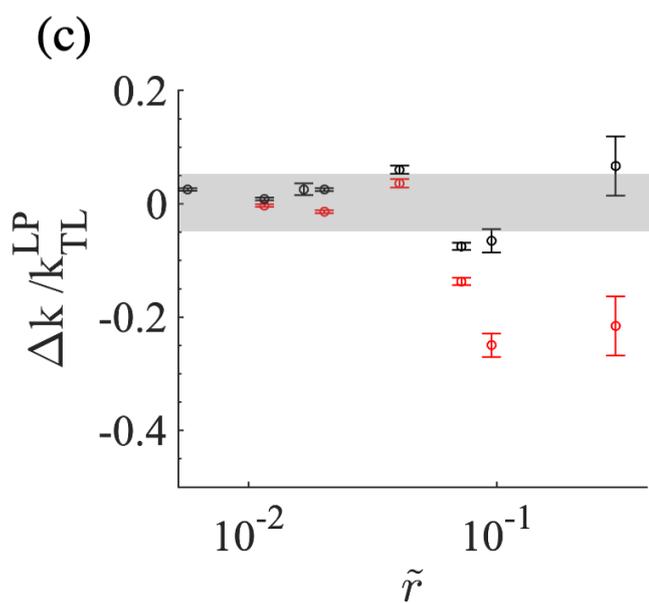

***Figure 8.*** *(a) Spring constant $k_{CP}^{LP}$ of CPs corrected according to Eq. 8 and Eq. 11 versus the reference spring constant $k_{TL}^{LP}$ of the TL cantilever calculated at the loading point (Eq. 8), measured before the sphere attachment. The continuous line represents the relation $k_{TL}^{LP} = k_{CP}^{LP}$. (b,c) The relative difference between $k_{CP}^{LP}$ and $k_{TL}^{LP}$ as a function of the reduced mass and radius of the CP, respectively, without (red crosses) and with (black circles) the correction by Eq. 11. The shaded area represents the ±5% interval.*

The direct calibration of the deflection sensitivity of large CPs by the standard method based on the acquisition of force curves can therefore be problematic. In this case, alternative approaches should be adopted. One possibility is to consider the mean of the inverse slopes of the loading and unloading portion of the force curve, as proposed by Chung et al.[71] The same authors proposed an electrostatic method to calibrate the deflection sensitivity, which however requires to build a dedicated micro-device and the metallisation of the CP[71]. In the following sections, we will propose an alternative method to overcome these limitations.

**4 BEST PRACTICES FOR THE CALIBRATION OF (LARGE) CPS**

In summary, the thermal noise method can be applied directly to the CP, following the following rule:

1. $S_z$ is measured carefully, neglecting the initial region of the FC after the contact point, and averaging the slopes of the loading and unloading curves to minimise friction-related effects.
2. The first resonant peak in the PSD is well sampled when applying the thermal noise method, to accurately estimate the power $P_V$, despite the typical peak sharpness in air. Alternatively, the thermal noise calibration can be carried out in water, in order to obtain a wider resonance peak. In this case, also the deflection sensitivity must be characterised in water.
3. The *β*-factor correction (Eq. 11) must be applied to the intrinsic spring constant calculated by Eq. 7. To this purpose, the sphere radius, the cantilever dimensions, and therefore the reduced mass $\tilde{m}$ and the reduced gyration radius $\tilde{r}$, must be accurately determined.
4. The corrected intrinsic spring constant is transformed into the apparent spring constant, including also the effect of the shift of the loading point, using Eq. 10.

According to the results reported in the present study, we suggest an alternative accurate and reliable approach for the calibration of CPs, and especially of large CPs. This approach can be applied when the CP is produced in house. In this case, the following procedure can be followed:

1. The intrinsic spring constant $k_{TL}$ of the TL cantilever, before the attachment of the sphere, is calibrated by the thermal noise method, measuring the deflection sensitivity in the standard way, i.e. from a FC acquired on a hard substrate, or using a deflection sensitivity -free approach, like LDV (and thermal noise)[51] or the Sader method[40,49,72].
2. Once the sphere is attached, measure the position of the loading point using an optical microscope and correct the intrinsic spring constant $k_{TL}$ into the apparent $k_{CP,app}^{LP}$, using Eq. 10.

It must be considered that, while the interferometric approaches can be used, together with Eq. 11, to calibrate CPs (in fact, LDV is a technique of choice for calibrating the spring constant of AFM probes, but at present it relies on expensive instrumentation), the applicability of the global calibration initiative launched by Sader[40] should be checked. This approach is based indeed on the statistical assessment of the hydrodynamic function for a given cantilever geometry and does not account for the contribution of the large attached sphere. The general Sader method for the calibration of cantilever of arbitrary shape and dimensions[73] could in turn work well, provided the calibration of the hydrodynamic function is performed in house.

Once the spring constant $k_{CP}^{LP}$ of a CP has been accurately calibrated, the deflection sensitivity $S_z$ of the current experiment can be calculated assuming that the thermal noise method, through Eq. 7, must provide the known value of the spring constant (see Ref[75] and the SNAP procedure[8]).

## 5 CONCLUSIONS

Colloidal probes are used in a wide range of experimental conditions. Their dynamical behaviour presents peculiar aspects[27], which also pose issues for their calibration, since the modal shapes are significantly affected.

In this work we have demonstrated that the addition of a large mass to a tipless cantilever concentrated at its end, besides significantly decreasing the resonant frequency, boosts the quality factor and the amplitude of the first resonant peak, so that a large CP mimics an SHO better than a standard AFM probe. These findings may have an impact on the application of CPs in spectroscopic dynamical applications.

Despite the relevant modifications of the cantilever dynamics of CPs, the accuracy of the calibration of CPs is not severely affected until the added mass of the sphere becomes comparable to that of the cantilever, which typically happens for diameters of the glass sphere of about 10 μm.

In the case of large CPs, it is possible to accurately calibrate the spring constant by applying a correction factor (Eq. 10) to the value obtained according to the equipartition theorem for an ideal tipless rectangular cantilever. We present two alternative detailed procedures, including one allowing to minimise the impact of the inaccurate determination of the deflection sensitivity.

## 6. ACKNOWLEDGMENTS

We acknowledge the support of the European Union's Horizon 2020 research and innovation programme under the Marie Skłodowska-Curie grant agreement No 812772, project Phys2BioMed, and under the FET Open grant agreement No. 801126, project EDIT. We thank Nadia Santo of the imaging facility noLimits (University of Milan) for the SEM images of the CPS.